\documentclass[multphys,vecphys]{svmult}

\usepackage{makeidx}     
\usepackage{graphicx}    
\usepackage{multicol}    

\usepackage{cite}
\usepackage{epsfig}

\makeindex             

\begin{document}

\title{The droplet evaporation$/$condensation transition in a finite volume}
\author{P. Virnau\inst{1}\and L.G. MacDowell\inst{2}\and M.M\"{u}ller\inst{1}\and K. Binder\inst{1}}
\institute{Institut f\"{u}r Physik, Johannes Gutenberg-Universit\"{a}t Mainz, Staudinger Weg 7, 55099 Mainz, Germany,
\texttt{Kurt.Binder@uni-mainz.de}
\and {Dpto. de Quimica Fisica, Univ. Complutense, Facultad de Cc. Qumicas, 28040 Madrid, Spain}}
%
%
\maketitle

A fluid in the $NVT$ ensemble at $T$ less than the critical
temperature $T_c$ and $\rho = N/V$ somewhat in excess of
$\rho ^{(1)} _{\textrm{coex}} $ (density of the saturated gas
in the gas-liquid transition) is considered. For $V \rightarrow
\infty $, a macroscopic liquid droplet coexists
with surrounding saturated gas according to the lever rule. 
For finite $V$, droplets can only exist
if they exceed a minimum size. A (rounded) first order
transition of the system occurs when the droplet evaporates into
the supersaturated gas.
Simulation evidence for this transition
is given for a Lennard-Jones model and interpreted by a phenomenological theory. 
At the transition, the chemical potential difference $\mu _t - \mu_{\textrm{coex}}$
scales like $L^{-d/(d+1)}$ for a cubic volume $V=L^d$ in $d$ dimensions, as $L \rightarrow \infty$.

\section{Introduction}
\label{sec:1}
There is much interest in the pathway of first order phase
transitions, where a droplet of the new phase is formed on the
background of the old phase \cite{1,2,3,4} by applying computer simulations
\cite{4,5,6,7,8,9}. For the liquid-gas transition, simulating $N$
particles interacting with a Lennard-Jones (LJ) potential in a
finite volume $V$ and applying periodic boundary conditions in
all directions, one finds a liquid droplet coexisting with
surrounding supersaturated gas for an appropriate range of $T$ and
$\rho$. From such data the excess free energy of the droplet and
the associated nucleation barrier can be inferred.

However, for all simulations \cite{10,11} one must be aware of
systematic errors due to finite size effects. 
Nucleation theory considers the (hypothetical) unstable equilibrium of one liquid
droplet coexisting with an infinite amount of surrounding
supersaturated gas. This saddle point in configuration space is
not accessible to a straight-forward simulation, though. For a
finite amount of surrounding supersaturated gas, the droplet is stable 
in a suitable parameter range \cite{6,12,13}. If we
increase the total volume (at fixed droplet volume), we encounter a
size where the droplet is no longer stable but rather evaporating
(with an increase in supersaturation
of the surrounding gas) \cite{6,12,13}.

Some aspects of this transition have recently been proven
rigorously \cite{13}. Implications are manyfold. The transition is e.g. 
crucial for the efficiency of multicanonical sampling \cite{14} 
in large systems. It can also help to understand
related phenomena such as the melting of islands adsorbed on
surfaces in the submonolayer range \cite{15,16} or microcanonical
simulations close to two-phase coexistence \cite{17}.

\section{Simulation evidence for the Lennard--Jones fluid}
\label{sec:2}
\begin{figure}
\begin{center}
\epsfig{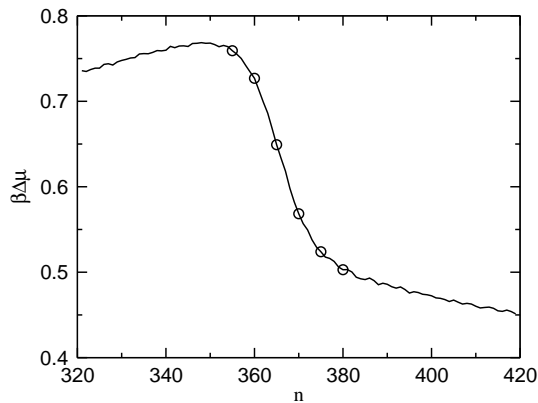}
\caption{\label{fig1}Plot of $\beta \Delta \mu$ vs. $N$, for a
$LJ$ fluid in a box of $L=22.5$ at $T=0.68$ 
($\Delta \mu = \mu_{NVT}-\mu_{\textrm{coex}}$). Note that lengths
are measured in units of $\sigma$, temperature in units of
$\varepsilon/k_B$, and $\beta=1/k_BT$.
$\circ$ denotes densities at which configurations were stored for analysis
(compare with Fig.~\ref{fig2})
}
\end{center}
\end{figure}

Grand canonical Monte Carlo simulations were carried out in $d=3$ dimensions using
a truncated and shifted LJ potential:
\begin{equation} \label{eq1}
U_{LJ} = 4 \epsilon \left [\left (\frac{\sigma}{r}\right)^{12} - \left(\frac{\sigma}{r^6}\right) + \frac{127}{16384} \right],  \hspace{0.1cm} \rm{if}
\; r \leq r_c = 2(2^{1/6} \sigma) \hspace{0.1cm} \rm{and} \hspace{0.1cm} 0 \hspace{0.1cm} \rm{else}.
\end{equation}
A multicanonical weighting scheme \cite{multican} was applied to simulate states
below $T_{c}$. Weighting functions were generated by adopting Wang-Landau
sampling \cite{Wang_Landau} to our off-lattice $(\mu VT)$ ensemble \cite{WL_LJ}.
From the resultant probability distribution $P_{\mu VT} (N)$ one can obtain
the chemical potential
\begin{equation} \label{eq2}
\mu _{NVT}= - kT \left( \frac{\partial \ln P_{\mu VT}(N)}{\partial N} \right)+\rm{const}
\end{equation}
and $\Delta \mu = \mu_{NVT}-\mu_{\textrm{coex}}$ with $\mu _{\textrm{coex}}=0$ being an estimate for
$\mu$ at the coexisting curve \cite{18}.
Inside the two-phase region and for finite $V$, $\Delta
\mu$ jumps to a lower value at an effective ``spinodal point'', 
which moves towards the coexistence curve as $V$ increases \cite{18} (Fig. \ref{fig1}).
\begin{figure}
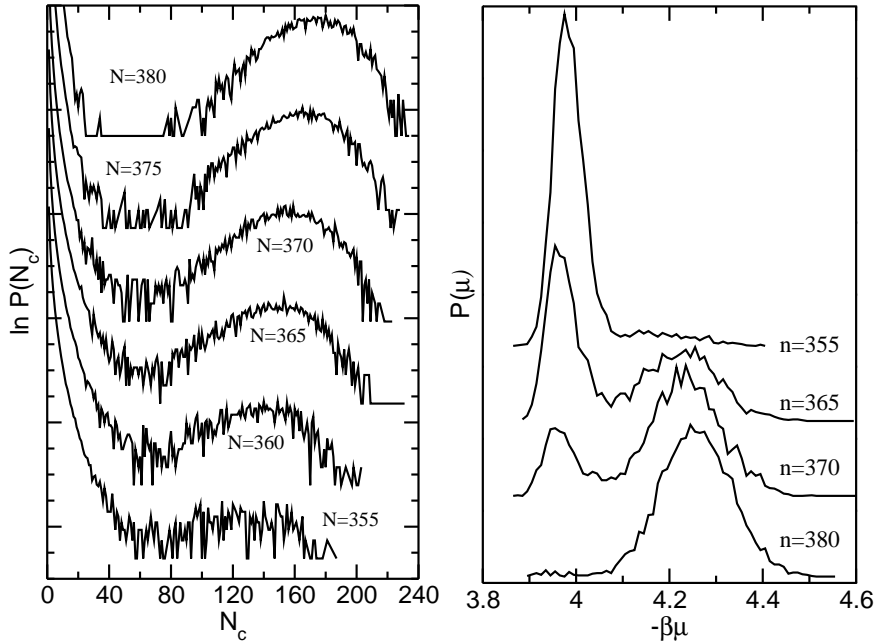

\epsfig{file=fig2a.eps,width=0.49\linewidth,clip=}
\epsfig{file=fig2b.eps,width=0.485\linewidth,clip=}
\caption{\label{fig2}Distribution $p(N_c)$ of the cluster size
$N_c$ for several choices of $N$ (left part) and the corresponding
distribution of the chemical potential of the supersaturated gas
$p(\mu)$ (right part). System parameters are the same as in
Fig.~\ref{fig1}.}
\end{figure}

While in principle this has been known for a long time \cite{19}, we now
present direct evidence from simulation. During the multicanonical run, we saved
configurations for typical densities (compare with Fig.~\ref{fig1}). After simulation,
the cluster size distribution $p(N_c)$ and the distribution $p(\mu)$ of 
the chemical potential $\mu$ in the gas region outside of the 
cluster was determined. (Fig.~\ref{fig2}). Any ensemble of atoms whose distance is smaller than
$1.5\sigma$ is assumed to belong to the same cluster \cite{20}. The chemical
potential was measured by the Widom particle insertion method \cite{21}. 
For $N \leq 350$, i.e. on the ascending branch of the $\Delta\mu $
vs. $N$ curve in Fig.~\ref{fig1}, $p(N_c)$ is monotonically
decreasing with $N_c$. For $N \approx 355$ a peak near $N_c \approx 120$ 
appears. It becomes more pronounced and moves to larger
$N_c$ as $N$ increases. This peak represents a single large liquid droplet,
present in the descending part of the $\Delta\mu $ vs. $N$ curve.
However, the liquid droplet cannot be found in all sampled
configurations: when we sample the distribution of the
largest cluster $N_c^{\textrm{max}}$ in the system \cite{18}, we
find a bimodal distribution with one peak near
$N_c^{\textrm{max}}\approx20$ in the whole range of $N$ studied in
Fig.~\ref{fig2}a. This peak corresponds to configurations which consist
of supersaturated gas with small clusters but no single large droplet.
The second peak is identical to the peak of the single droplet as shown in Fig.~\ref{fig2}a.
This clear cut evidence for the ``evaporation'' of the liquid droplet also appears in double peak
- distributions of the chemical potential (Fig.~\ref{fig2}b) and
the internal energy \cite{18}.
\section{Phemonenological theory using the Ising lattice gas model and concluding remarks}
\label{sec:3}
The discussion is simplest for the lattice gas, due to its perfect
symmetry between liquid and gas, brought out in the magnetic
interpretation: the two coexisting phases have spontaneous
magnetization $\pm m_\textrm{coex}$, and $\Delta\mu$ corresponds
to the magnetic field. The thermodynamic potential
$g_{\textrm{s.g}}(n)$ of the supersaturated gas is simply expanded
at the coexistence curve, for magnetization $m$ near
--$m_{\textrm{coex}}$:
\begin{equation} \label{eq3}
g_{\textrm{s.g}} (m) =g_0(T) + \frac{1}{2 \chi_{\textrm{coex}}} (m
+ m_{\textrm{coex}})^{2},
\end{equation}
$\chi_{\textrm{coex}}$ being the susceptibility (per lattice site)
at the coexistence curve. Similarly, the state containing a
droplet is described by
\begin{equation} \label{eq4}
g_{\textrm{drop}} (m) =g_0 (T) + \frac{1}{2 \chi_{\textrm{coex}}}
(m'+ m_{\textrm{coex}})^2 + \frac{S_d \gamma (R^{*})^{d-1}}{L^d},
\end{equation}
where the last term describes the surface free energy of a droplet
(per lattice site) in the capillarity approximation, $S_d$ being
the surface area of a $d$-dimensional unit sphere, $\gamma$ the
interfacial tension, and $R^*$ the critical droplet radius
(satisfying $R^*=(d-1) S_d \gamma / [2 dm_{\textrm{coex}} \quad
V_d H]$ $V_d$ being the volume of the unit sphere), and $m'
+m_{\textrm{coex}} = \chi_{\textrm{coex}} H$ for $H \rightarrow
0$.
The transition occurs for $g_\textrm{s.g} (m)=g_{\textrm{drop}}
(m)$ \cite{12}. Rescaling variables $\widetilde{m}=m/
m_{\textrm{coex}}$, $\widetilde{H}=H
\chi_{\textrm{coex}}/m_\textrm{coex}$ one finds \cite{13} a first
order transition at $\widetilde{m}_t=-1 + (d+1) (S_d/2d)^{d/(d+1)}
V_d ^{-(d-1)/(d+1)} c^{d/(d+1)} \, \widetilde{L}^{-d/(d+1)}$,
where the field exhibits a jump between $\widetilde{H}^{(1)}_t=1 +
\widetilde {m}_t$ and $\widetilde {H}^{(2)}_t=(d-1)
\widetilde{H}^{(1)}_t / (d+1)$,
$\widetilde{L}=L/\xi_{\textrm{coex}}$, $\xi_{\textrm{coex}}$ being
the correlation length, and $c$ is a constant that is universal
near $T_c \{c=\gamma \chi_{\textrm{coex}} /
m^2_{\textrm{coex}}\xi_{\textrm{coex}}$ \cite{12}\}. The variation
with $L^{-d/(d+1)}$ was first noted in \cite{6} and recently
proven rigorously \cite{13}. Of course, for finite $L$ this
anomalous first order transition is rounded, but it becomes sharp
as $L \rightarrow \infty$ \cite{13}. 


\section*{Acknowledgements} 
Financial support from the BASF AG (P.V.) and from the Deutsche For\-schungsgemeinschaft (DFG) via a
Heisenberg fellowship (M.M.) and by grant No BT 314/17-3 (L.G.M) is gratefully acknowledged.
LGM would like to thank Ministerio de
Ciencia y Tecnologia (MCYT) and Universidad Complutense 
(UCM) for the award of a Ramon y Cajal fellowship and for
financial support under contract BFM-2001-1420-C02-01.
We are also grateful to 
NIC J\"ulich, HLR Stuttgart, and ZDV Mainz for generous grants of computer time.

%
%



\printindex
\end{document}